\documentclass[
  ,draft            
  ]
  {aipproc}

\layoutstyle{6x9}

\newcommand{\la}{\langle}
\newcommand{\ra}{\rangle}

\newcommand{\beq}{\begin{eqnarray}}
\newcommand{\eeq}{\end{eqnarray}}
\newcommand{\sbeq}{\begin{subeqnarray}}
\newcommand{\seeq}{\end{subeqnarray}}

\newcommand{\bhatr}{\mbox{{\boldmath ${\hat r}$}}}

\newcommand{\bfx}{\mbox{{\boldmath $x$}}}

\newcommand{\btau}{\mbox{{\boldmath $\tau$}}}
\newcommand{\bsigma}{\mbox{{\boldmath $\sigma$}}}

\begin{document}

\title{Chiral Symmetry and Axial Anomaly in Hadron and Nuclear Physics --- a review ---}


\author{Teiji Kunihiro}{
  address={Department of Physics, Kyoto University, Kitashirakawa, Sakyoku,
 Kyoto 606-8502, Japan}
}



\begin{abstract}
The important role played by chiral symmetry and axial anomaly in QCD in nuclear 
physics is reviewed. Some recent topics on possible chiral restoration in hot and/or
 dense matter are picked up. We also discuss
 so called {\em effective} restoration of chiral 
anomaly hot and/or dense matter, 
as may be seen in a character change of  $\eta '$ meson.
\end{abstract}

\maketitle


\section{Introduction}

Nuclear physics was primarily  quantum many-body physics with the nuclear force given,
which is responsible for binding the nucleons against the 
repulsive Coulomb force between protons.
Yukawa's meson theory\citep{Yukawa:1935xg}
 was the first intended application of 
quantum-field theory to the problem of  the nuclear 
force. The salient ingredients of the nuclear force\citep{tamagaki68} are 
the tensor force\citep{OPEP} in
the long range and the short-range repulsive core\citep{jastrow}.
The tensor force is generated by
 one-pion exchange between the nucleons.
The one-pion-exchange potential (OPEP)\citep{OPEP} reads
\beq
V_{\rm OPEP}(1,2)=
f^2m_{\pi}\frac{\btau _1\cdot\btau _2}{3}\Bigl[(\bsigma _1\cdot\bsigma _2)Y(m_{\pi}r)+S_{12}Z(m_{pi}r)\Bigl],
\eeq
where $Y(x)={\rm exp}(-x)/x)$,
 $Z(x)=(1+3/x+3/x^2)Y(x)$ and 
$S_{12}= 3(\bsigma_1\cdot\bhatr)\bsigma_2\cdot\bhatr)-
\sigma_1\cdot\bsigma_2 = 3\sqrt{5}\big[\bsigma_1\otimes\bsigma_2]^{(2)}
\otimes [\bhatr_1\otimes\bhatr_2]^{(2)}\big]^{(0)}$
being the tensor operator which is constructed from 
the two second-rank tensors. The appearance of such an operator with a bad symmetry
is due to the fact that the pion is a pseudo-scalar particle.

Owing to the transformation properties of the tensor force,
it only acts to the spin-triplet state but not to the singlet state, 
which is the reason why deuteron exists as a proton-neutron bound
system although there are no di-neutron bound system: The second order
contribution of the tensor force gives rise to an additional
attraction between the triplet state.
This second-order effect of the tensor force is also an essential
ingredient for realizing the saturation property of the nuclear matter\citep{saturation}.

Then why does pion is isovevtor and pseudo-scalar particle with the
lightest mass in the hadron world?
These are all because the pion is the Nambu-Goldstone boson associated
with dynamical breaking of chiral symmetry of QCD\citep{Nambu:1961tp}.
How important roles does the chiral symmetry play 
in nuclear physics? Some answers may be found in 
\citep{Hatsuda:1994pi,Hatsuda:2001da}.
Before answering this problem, we clarify the chiral symmetry  and
its spontaneous breaking in QCD.

\section{Chiral invariance of classical QCD Lagrangian}

The classical QCD Lagrangian reads
\beq
{\cal L}=\bar{q}(i\gamma^{\mu}D_{\mu}-m)q-\frac{1}{4}F^a_{\mu\nu}F_a^{\mu\nu}.
\eeq
The classical QCD Lagrangian
with vanishing quark mass ($m \rightarrow 0$) is invariant
under the chiral transformation.
The chiral transformation for $N_F$-flavor quark field $q_{f}$
($f = 1,\, 2,\, ,,, N_F$)  is defined as a direct product of 
two unitary transformations $U_L$ and $U_R$;
\beq
q_{L\, f}\equiv \frac{1-\gamma_5}{2}\, q_{f} \rightarrow (U_L)_{ff'}q_{L\, f'},\quad
q_{R\, f}\equiv \frac{1+\gamma_5}{2}\, q_{f} \rightarrow (U_R)_{ff'}q_{R\, f'}.
\eeq
Notice that  the vector current
$\bar{q}\gamma^{\mu}q=\bar{q}_L\gamma^{\mu}q_L+
\bar{q}_R\gamma^{\mu}q_R$ is invariant under the chiral
transformation, although the Dirac mass term $\bar{q}q=\bar{q}_R
q_L+\bar{q}_Lq_R$ is not.
If we neglect the current quark mass term,
the quark filed enters QCD only as a combination
$\bar{q}\gamma^{\mu}D_{\mu}q$, and hence it become invariant under
the chiral transformation. A warning is in order here; the axial $U(1)$ symmetry
is explicitly broken by a quantum effect, which is known as $U(1)_A$ 
anomaly\citep{Weinberg:1996kr}.

A quark bilinear operator $\Phi_{ij}$ defined by
$\Phi_{ij}=\bar{q}_j(1-\gamma_5)q_i=2\bar{q}_{j\,R}q_{i\, L}$
is  transformed as follows,
\beq
\Phi_{ij}\, \rightarrow \,(U_{L})_{ik}\Phi_{lk}(U_{R}^{\dag})_{lj}.
\eeq

In the two-flavor  case, the generators of the chiral transformation 
are given by the isospin charges $Q^a$ and the axial charges
$Q^a_5$;
\beq
Q^a=\int d\bfx \bar {q}\gamma^0\tau^{a}q/2,\quad
Q^{a}_5=\int d\bfx \bar {q}\gamma^0\gamma_5\tau^{a}q/2.
\eeq
We note the commutation relation,
\beq
[iQ^a_5, \bar{q}i\gamma_5\tau^bq]=-\delta^{ab}\bar{q}q.
\label{ccr}
\eeq
Then taking the vacuum expectation value of (\ref{ccr}),
we have
\beq
\la 0\vert \bar{q}q\vert 0\ra
=\la 0\vert [Q^a_5, \bar{q}\gamma_5\tau^aq]\vert 0\ra,
\eeq
which implies that 
if $\la 0\vert \bar{q}q\vert 0\ra\not= 0$,
then $Q^a_5\vert 0\ra$ can not be zero for some $a$.
That is, chiral symmetry is spontaneously broken!
Indeed there is a following celebrated relation 
due to  Gell-Man, Oakes and Renner\citep{GellMann:1968rz},
\beq
f_{\pi}^2m_{\pi}^2 = -\frac{m_u+m_d}{2}\la 0\vert \bar{u}u+\bar{d}d\vert 0\ra,
\eeq
which does indicate that the chiral symmetry is spontaneously broken in 
the QCD vacuum, because the pion decay constant $f_{\pi}\simeq 93$ MeV
is finite.

\section{Possible chiral restoration in finite nuclei}

One of the interesting nature of QCD is that the QCD vacuum can
change along with an inclusion of external hard scale, which may be induced by
baryon chemical potential,i.e., the baryon density, temperature, strong 
magnetic field and so on. 
An interesting observation is that a nucleus can provide 
a hard scale by its baryon density, which might cause a change of
the QCD vacuum, and hence the chiral symmetry may be 
partially restored in a finite nucleus.
Thus exploring possible evidence of partial restoration of chiral symmetry in 
the nuclear medium has become  one of the 
most important and challenging problems 
in nuclear physics\citep{Hatsuda:2001da,Weise:2008bk}.  
Relevant experimental studies include
 the spectroscopy of deeply bound pionic atoms \citep{Suzuki:2002ae},
 low energy pion-nucleus scatterings  \citep{Friedman:2004jh}, 
and the production of di-pions  in hadron-nucleus and photon-nucleus
reactions \citep{Bonutti:2000bv,Starostin:2000cb,Messchendorp:2002au}.
 These experiments revealed the following anomalous properties of the pion dynamics
in the nuclear medium;
(i)~  an enhancement of the repulsion  
$\pi^{-}$-nucleon interaction\citep{Suzuki:2002ae,Friedman:2004jh},
(ii) an enhanced attraction of the  $\pi$-$\pi$ interaction in the
scalar-isoscalar channel\citep{Bonutti:2000bv,Starostin:2000cb,Messchendorp:2002au}.

In the theoretical side,
possible relevance of the $\pi$-$\pi$ interaction in a nuclear medium
was first suggested in \citep{Hatsuda:1999kd}.
Weise and his collaborators showed that that the reduction of the 
temporal part of the pion decay constant in the nuclear medium
$F_{\pi}^t$ is intimately related to the anomalous repulsion (i)
 \citep{Kolomeitsev:2002gc,Weise:2005ss}. 
It was also argued that the reduction of $F_{\pi}^t$ is responsible for 
the phenomenon (ii) \citep{Jido:2000bw}.

Recently, Jido, Hatsuda and the present author\citep{Jido:2008bk}
 derived a novel sum rule for the quark condensate valid for all density, 
which sum rule is reduced  to 
\beq
\langle \bar{q}q \rangle^*/\langle \bar{q}q \rangle =
(F_{\pi}^t/F_{\pi}) Z_{\pi}^{*1/2}
\eeq
in the low-density limit. Here
 $\langle \bar{q}q \rangle^*$ is the quark condensate,
$F_{\pi}^t$ the (temporal) pion decay constant  
and the pion wave-function renormalization constant
$Z_{\pi}^{*}$ all in the nuclear medium.
It is noteworthy that the $Z_{\pi}^{*}$ can be estimated
with the use of the 
the iso-singlet pion-nucleon scattering amplitude at low energy,
and they found that 
\beq
{Z_{\pi}^{*}}^{1/2}=\left(\frac{G_{\pi}^*}{G_{\pi}}\right)^{1/2}=1-\gamma \frac{\rho}{\rho_0},
\eeq
where $G_{\pi}^{(*)}$ is the (in-medium) pion coupling constant.
Here the coefficient $\gamma=\beta \rho_0/2=0.184$ with
$\beta = 2.17\pm 0.04 $fm$^3$.
(Parametrically, $\beta$ is expressed as
\beq
\beta=\frac{\sigma _{\pi N}}{F_{\pi}^2m_{\pi}^2}+
   \left(1+\frac{m_{\pi}}{m_{N}}\right)\frac{4\pi a_{\pi
   N}}{m_{\pi}^2}.
\eeq
Here $\sigma_{\pi N}$ and $a_{\pi N}$ are the $\pi$-N sigma term
and the iso-singlet scattering length, respectively.)
Then the in-medium quark condensate is nicely expressed 
by the temporal pion decay constant and the pion coupling;
\beq
\la \bar{q}q\ra^*=- F_{\pi}^t G_{\pi}^{*}.
\eeq

Now the $s$-wave $\pi^{-}$-nucleus optical potential $U_s$ is 
parametrized as
\beq
2m_{\pi}U_s &=& -4\pi
[1+\frac{m_{\pi}}{m_N}](b_0^{*}\rho_{+}-b_1^{*}\rho_{-}), \nonumber \\
  &=& -T^{(+)*}(\omega=m_{\pi}:m_{\pi})\rho_{+}-
T^{(-)*}(\omega=m_{\pi}:m_{\pi})\rho_{-},
\eeq
where $\rho_{\pm}=\rho_p\pm\rho_n$ with $\rho_{p(n)}$ being the
proton (neutron) density. Then the parameter $b_1^{*}$ which can be
extracted from the experimental data is expressed as
\beq
\frac{b_1}{b_1^{*}}=\left(\frac{F_{\pi}^t}{F_{\pi}}\right)^2.
\eeq

Combining these relations, Jido et al\citep{Jido:2008bk} derived
the following relation
\beq
\frac{\la\bar{q}q\ra^{*}}{\la \bar{q}q\ra}\simeq \left(\frac{b_1}{b_1^*}\right)^{1/2}
\left(1-\gamma \frac{\rho}{\rho_0}\right).
\eeq
Now one sees that the experimental evidence of the repulsive enhancement as
 given by $b_{1}^{*}$ implies that the absolute value of the 
quark condensate in the nuclei is smaller than that in the vacuum, and hence the
chiral symmetry is partially restored in the nuclei.

\section{{\em Effective} restoration of axial symmetry at finite temperature
 and density}

How about other signals of the 
chiral restoration at finite density and/or temperature?
The bottom line is that some  hadrons are intimately related to 
the chiral symmetry and its dynamical breaking, and hence their
properties may change along 
with the chiral transition at finite density/temperature\citep{Hatsuda:1994pi}.
Such hadrons include the sigma meson\citep{Hatsuda:1986gu}.
The chiral symmetry implies that the degeneracy of the vector and axial vector
correlators as well as that in the scalar and pseudoscalar channels,
which are parity partners.
Thus exploring the possible tendency of the degeneracy in these opsite
parity channels
should be interesting in an extreme environment.
The parity doubling in the baryon sector may be affected by the 
underlying chiral symmetry\citep{Detar:1988kn}. 
Examining the properties of the negative-parity
baryons such as $N^*(1535)$ at finite density and/or temperature should be also
interesting\citep{Nagahiro:2003iv}.
An interesting ingredient in this subject lies in the fact that
$N^*(1535)$ is strongly coupled with $\eta$ meson.
Thus the study of $N^{*}(1535)$ is automatically to explore the
properties of $\eta$ meson in nuclei. One should also note that
$\eta$ meson is a mixing partner of $\eta'(958)$, the nature of which
is intimately related with the axial anomaly of QCD\citep{Nagahiro:2004qz}.

One of the fundamental properties of QCD is 
$U(1)_A$ anomaly or axial anomaly\citep{Weinberg:1996kr}.
The ninth pseudoscalar meson $\eta'$ which is almost flavor singlet 
with a mass as large as 958 MeV
is a reflection of the $U(1)_A$ anomaly and the $\theta$ vacuum
owing to the instanton configuration.
The mass of $\eta$ and $\eta'$ and their mixing property 
are realized with combined effects of the anomaly, explicit and dynamical
breaking of chiral symmetry.
The instanton density as well as the quark condensates is expected to
decrease at
 finite temperature and density\citep{Gross:1980br,Schafer:1996wv}.
Thus one should also explore the properties of the $\eta$-$\eta'$ meson sector
at finite temperature and/or density\citep{Kunihiro:1989my},
 which may show an {\em effective} restoration
of $U(1)_A$ symmetry\citep{Pisarski:1983ms,Kunihiro:1989my,Schafer:1996wv} 
as seen in the mixing properties of $\eta$ and $\eta'$ mesons 
and their masses.

The $U(1)_A$ anomaly implies that the $U(1)_A$ symmetry is 
explicitly broken by a quantum effect. In the context of the effective
Lagrangian, there
should exist a vertex which violates this symmetry. One of such an
interaction 
is the six-quark interaction with 
a determinantal form as introduced by Kobayshi and
Maskawa\citep{Kobayashi:1970ji}
in 1970,
\beq
{\cal L}_{KMT}=g_{_D}\det_{i,j}\bar{q}_i (1-\gamma_5) q_j + {\rm h.c.},
\label{eq:kmt}
\eeq
where h.c. stands for Hermite conjugate.
This vertex is contained in instanton-induced quark interaction
derived
by 't Hooft in 1976\cite{'tHooft:1976fv}; see \citep{Hatsuda:1994pi}
for a review.
It was first shown by the present author \cite{Kunihiro:1989my} 
using a generaized Nambu-Jona-Lasinio model incorporating
the Kobayashi-Maskawa-'t Hooft term (\ref{eq:kmt})
 that 
the $\eta$ and $\eta^\prime$ mesons change
their nature owing to both the temperature dependence of
 the quark condensates and the possible decrease
in the KMT coupling constant $g_{_D}$ with $T$.
The coupling constant $g_{_D}$ of the KMT term may be dependent 
on temperature and baryon chemical potential because the instanton
density is
dependent on them\cite{Gross:1980br,Schafer:1996wv}.
Such a possible temperature dependence causes
a temperature dependence of the mixing angle $\theta_{\eta}$ so that 
  $\theta_{\eta}$  increases in the absolute value and
the mixing between the $\eta$ and $\eta'$ approaches the ideal 
one.
Although the $\eta_0$ component in the physical $\eta'$ 
decreases as $T$ is increased,
the $\eta^{\prime}$  mass decreases gradually with increasing
$T$, because 
the $\eta_0$ tends to acquire the nature
of the ninth Nambu-Goldstone boson of
the $SU(3)_L \otimes SU(3)_R \otimes U(1)_A$ symmetry 
and decreases its mass rapidly. This is an effective restoration 
$U(1)_A$ symmetry first discussed by Pisarski and Wilczek\citep{Pisarski:1983ms}. 
Such an anomalous decrease in the $\eta^{\prime}$ mass
might have been observed in 
the relativistic heavy ion collisions at RHIC\cite{csorge}.
 Recent studies on this problem are reviewed in 
\citep{Kunihiro:2009ds}.

\section{Brief summary}

The following is a  summary what I wanted to say in this report:
(1)~The saturation property of the
nuclear matter can be attributed eventually to chiral symmetry and its 
dynamical breaking in QCD.
(2)~Hadrons are sort of elementary excitations on top of
the nonperturbative QCD vacuum, and hence 
may change their properties along with that of the QCD vacuum.
(3)~The QCD vacuum can change and even show a phase transition(s)
with an increase of  temperature and/or baryon density,
 which in turn gives rise to a
 change of particle pictures of the hadrons in the system.

\begin{theacknowledgments}
I thank the organizers , in particular, Professor Ozawa
 to invite me to this interesting workshop.
This work was partially supported by a Grant-in-Aid for Scientific
Research by the Ministry of Education, Culture, Sports, Science
and Technology (MEXT) of Japan (No. 20540265)
 and by the
Grant-in-Aid for the global COE program `` The Next Generation of
Physics, Spun from Universality and Emergence '' from MEXT.

\end{theacknowledgments}




\begin{thebibliography}{99}
\bibitem{Yukawa:1935xg}
  H.~Yukawa,
  Proc.\ Phys.\ Math.\ Soc.\ Jap.\  {\bf 17}, 48 (1935)  .
%
\bibitem{tamagaki68}
R. Tamagaki, Prog. Theor. Phys. {\bf 39}, 91 (1968); \\
R. V. Reid, Ann. of Phys. {\bf 50}, 411 (1968) ;\\
 M. Taketani, R. Tamagaki, W. Watari, S. Machida, S. Ogawa, T. Ueda,
W. Watari, M. Yonezawa, S. Furuichi and K. Nisimura,
Prog. Theor. Phys. Suppl.  {\bf 39} (1967).\\
 N. Hoshizaki and S. Otsuki, Prog. Theor. Phys. Suppl. {\bf 42}
(1968).

%
\bibitem{OPEP}
M. Taketani, J. Iwadare, S. Otsuki, R. Tamagaki, S. Machida,
T. Toyoda, W. Watari and K. Nishijima, Prog. Theor. Phys. Suppl. {\bf 3}
(1956).
%
\bibitem{jastrow}
R.~ Jastrow, Phys. Rev. {\bf 81},  165 (1950).
%
\bibitem{saturation} As review articles, see, for example, H.~A.~ Bethe, R
Annu. Rev. Nucl. Sci. {\bf 21}, 93 (1971) ; \\
 B. Day, Rev. Mod. Phys. {\bf 50} , 495 (1978).
%
\bibitem{Nambu:1961tp}
  Y.~Nambu and G.~Jona-Lasinio,
  Phys.\ Rev.\  {\bf 122}, 345 (1961).
%
\bibitem{Hatsuda:1994pi}
  T.~Hatsuda and T.~Kunihiro,
  Phys.\ Rept.\  {\bf 247}, 221 (1994)
%
\bibitem{Hatsuda:2001da}
  T.~Hatsuda and T.~Kunihiro,
  arXiv:nucl-th/0112027.
%
\bibitem{Weinberg:1996kr}
 S.~Weinberg,
{\it The quantum theory of fields. Vol. 2: Modern applications}
   (Cambridge University Press, UK, (1996);\\
%
  K.~Fujikawa and H.~Suzuki,
{\it Path integrals and quantum anomalies}
  Oxford, UK: Clarendon, (2004).
%
\bibitem{GellMann:1968rz}
  M.~Gell-Mann, R.~J.~Oakes and B.~Renner,
  Phys.\ Rev.\  {\bf 175}, 2195 (1968).
%
\bibitem{Weise:2008bk}
  W.~Weise,
  Nucl.\ Phys.\  A {\bf 805}, 115 (2008)
  [arXiv:0801.1619 [nucl-th]].
%
\bibitem{Suzuki:2002ae}
K.~Suzuki {\em et~al.},
 Phys. Rev. Lett. {\bf 92}, 072302 (2004) ;
 P.~Kienle and T.~Yamazaki,
 Prog.\ Part.\ Nucl.\ Phys.\  {\bf 52} ,  85 (2004).

\bibitem{Friedman:2004jh}
E.~Friedman {\em et~al.},
Phys. Rev. Lett. {\bf 93}, 122302 (2004)
Phys.\ Rev.\ C {\bf 72} , 034609 (2005) .

\bibitem{Bonutti:2000bv}
F.~Bonutti {\em et al.}  [CHAOS collaboration],
Phys. Rev. Lett. {\bf 77},  603 (1996);
Nucl.\ Phys.\ A {\bf 677},  213 (2000);
P.~Camerini {\em et~al.}  [CHAOS collaboration], 
Nucl.\ Phys.\ A {\bf 735}, 89 (2004) .


\bibitem{Starostin:2000cb}
A.~Starostin {\em et~al.} [Crystal Ball Coll.], 
Phys. Rev. Lett. {\bf 85}, 5539 (2000) ;
Phys.\ Rev.\ C {\bf 66}, 055205 (2002) .

\bibitem{Messchendorp:2002au}
J.~G.~Messchendorp {\em et al.},
Phys.\ Rev.\ Lett.\  {\bf 89},  222302 (2002).
%
\bibitem{Hatsuda:1999kd}
  T.~Hatsuda, T.~Kunihiro and H.~Shimizu,
  Phys.\ Rev.\ Lett.\  {\bf 82}, 2840 (1999).

\bibitem{Kolomeitsev:2002gc}
E.~E. Kolomeitsev, N.~Kaiser, and W.~Weise,
 Phys. Rev. Lett. {\bf 90}, 092501 (2003) .

\bibitem{Weise:2005ss}
W.~Weise,
arXiv:nucl-th/0507058.

\bibitem{Jido:2000bw}
D.~Jido, T.~Hatsuda, and T.~Kunihiro,
 Phys. Rev. {\bf D63},011901 (2001) . 


\bibitem{Jido:2008bk}
  D.~Jido, T.~Hatsuda and T.~Kunihiro,
  Phys.\ Lett.\  B {\bf 670}, 109 (2008).
See also 
D.~Jido, T.~Hatsuda and T.~Kunihiro,
Prog.\ Theor.\ Phys.\ Suppl.\  {\bf 168} ,  478 (2007)
[arXiv:0706.0258 [nucl-th]].
%
\bibitem{Hatsuda:1986gu}
 T.~Hatsuda and T.~Kunihiro,
  Prog.\ Theor.\ Phys.\  {\bf 74}, 765 (1985);\\
  Phys.\ Lett.\  B {\bf 185}, 304 (1987).
%
\bibitem{Detar:1988kn}
  C.~E.~Detar and T.~Kunihiro,
  Phys.\ Rev.\  D {\bf 39}, 2805 (1989).
%
\bibitem{Nagahiro:2003iv}
  H.~Nagahiro, D.~Jido and S.~Hirenzaki,
  Phys.\ Rev.\  C {\bf 68}, 035205 (2003).
%
\bibitem{Nagahiro:2004qz}
  H.~Nagahiro and S.~Hirenzaki,
  Phys.\ Rev.\ Lett.\  {\bf 94} , 232503 (2005).

%
%
\bibitem{Gross:1980br}
  D.~J.~Gross, R.~D.~Pisarski and L.~G.~Yaffe,
  Rev.\ Mod.\ Phys.\  {\bf 53} , 43 (1981).
%
%
\bibitem{Schafer:1996wv}
  T.~Schafer and E.~V.~Shuryak,
  Rev.\ Mod.\ Phys.\  {\bf 70} , 323 (1998).
%
\bibitem{Kunihiro:1989my}
  T.~Kunihiro,
  Phys.\ Lett.\  B {\bf 219}, 363 (1989); \\
 Nucl.\ Phys.\  B {\bf 351}, 593 (1991).
%
\bibitem{Pisarski:1983ms}
  R.~D.~Pisarski and F.~Wilczek,
  Phys.\ Rev.\  D {\bf 29} , 338 (1984).
%
\bibitem{Kobayashi:1970ji}
  M.~Kobayashi and T.~Maskawa,
  Prog.\ Theor.\ Phys.\  {\bf 44} , 1422 (1970);\\
  M.~Kobayashi, H.~Kondo and T.~Maskawa,
  Prog.\ Theor.\ Phys.\  {\bf 45} , 1955 (1971).
%
\bibitem{'tHooft:1976fv}
  G.~'t Hooft,
  Phys.\ Rev.\  D {\bf 14} , 3432 (1976)
  [Errata; {\bf 18} , 2199 (1978)];\,
Phys.\ Rep.\  {\bf 142} , 357 (1986).
%
%
\bibitem{csorge}
R.~ V\'ertesi, T.~ Cs\"org\H{o} and J.~ Sziklai,
 Nucl.\ Phys.\  A {\bf 830}, 631C (2009):\,
arXiv:0905.2803.
%
\bibitem{Kunihiro:2009ds}
Roles of  the $U(1)_A$ anomaly at finite temperature and/or density 
is reviewed in \\
  T.~Kunihiro,
  Prog.\ Theor.\ Phys.\  {\bf 122}, 255 (2009).
  [arXiv:0907.3808 [hep-ph]].
%

\end{thebibliography}


\end{document}